\documentstyle[aps,epsfig,multicol]{revtex}
\begin{document}
\draft

\title{
Quantum Transparency of Barriers for Structure Particles
 }
\author{F.M.Pen'kov}
\address{Joint Institute for Nuclear Research,\\
141980, Dubna, Russia\\
penkov@thsun1.jinr.dubna.su}
\date{\today}
\maketitle

\begin{abstract}
Penetration of two coupled particles through a repulsive barrier is
considered. A simple mechanism of the appearance of barrier resonances is
demonstrated that makes the barrier anomalously transparent as compared to
the probability of penetration of structureless objects. It is indicated that
the probabilities of tunnelling of two interacting particles from a false
vacuum can be considerably larger than it was assumed earlier.
\end{abstract}

\pacs{PACS number(s): 11.10.Jj, 03.65.Nk, 21.45.+v}
\begin{multicols}{2}

\section{Introduction}
Quantum tunnelling through a barrier is one of the most common problems in
many trends of physics. Physical processes of tunnelling are usually
considered as a penetration of a structureless particle
through the barrier; whereas in realistic physics, we are, as a rule, dealing
with the problem of penetration of structure particles through the barrier.
It is clear that when the spatial size of a barrier is much larger than
the typical dimension of an incident complex, we could expect insignificant
distinctions of the penetration probability of the complex from that of
structureless particles. The situation drastically changes when the size of
the complex is larger than the spatial width of the barrier. In this case,
there arise mechanisms that increase the barrier transparency (see, for
instance, ref. \cite{Zakh} and references therein). The simplest of them arises
when only part of the complex interacts with the barrier, i.e. the
penetration probability depends on the mass smaller than the complex mass.

In this paper, we consider the mechanism of drastic transparency of a barrier
that implies possible formation of a barrier resonance. To this end, it is
necessary that at least two particles interact with the barrier. It is easy
to imagine the mechanism of formation of such a resonance state. Let through
the barrier, only one of particles pass, and forces coupling the pair be
sufficient to keep the particles on different sides of the barrier. Then,
such a resonance state will live unless one of the particles penetrates
through the barrier. The barrier width will determine the lifetime of a
resonance of that sort. As will be shown below, the probability of tunnelling
through the barrier can reach unity. Physically, this effect is explained
by the interference suppression of a reflected wave, since the presence
of a barrier resonance is simply described by the effective interaction in
the variable of motion of the centre of inertia of the pair, whose spatial
form has a local minimum in the barrier centre. Therefore, the suppression
of a reflected wave can be explained by the interference phenomenon well
known in optics and used in coating lenses -- the difference in path
between the wave reflected from the first peak and the one reflected from
the other peak should equal one-half the wave length.

In this paper, the
above effect of transparency is demonstrated analytically and numerically
for a pair of identical particles coupled by an oscillator interaction
(in what follows, an oscillator) that penetrates through a one-dimensional
repulsive barrier of the Gaussian type. This choice of interactions is,
on the one hand, due to the system being extremely simple and allowing  the
reduction of three-dimensional scattering of a three-dimensional pair of
particles on a one-dimensional barrier to the scattering of a one-dimensional
oscillator on a one-dimensional barrier. On the other hand, just this
type of interaction is taken in the literature~\cite{Rub1} devoted to the
decay probability of false vacuum  in the high energy
particle collisions (see, for instance~\cite{Rub2,Tyn1}). It was pointed out
that
the processes of transition from the false vacuum could be described on the
basis of quantum-mechanical tunnelling of a pair of particles through a
barrier, but a system was investigated, in which only one particle  of the
oscillator interacted with the barrier. In this note it is shown that when
two particles interact with the barrier, the penetration probability can
be essentially higher that in the systems considered earlier.

\section{Equations}
Consider the penetration of a pair of identical particles with masses
$m_1=m_2=m$
and coordinates  $ {\bf r}_1$ and  $ {\bf r_2}$ coupled by an oscillator
through the potential barrier  $ V_0(x_1)+V_0(x_2)$.
The Hamiltonian of this system ($\hbar $ = 1)
 $$
 -\frac{1}{4m}\triangle_R -\frac{1}{m}\triangle_r + \frac{m\omega^2}{4} r^2 +
 V_0({\bf R-r/2}) + V_0({\bf R+r/2}),
 $$
written in terms of coordinates of the centre of inertia of the pair
${\bf R} = ({\bf r}_1 + {\bf r}_2)/2 $
and internal variable of the relative motion
$ {\bf r}= {\bf r}_1 -{\bf r}_2 $
describes the three-dimensional motion of a three-dimensional oscillator.
Since the potential
barrier depends only on one variable, and the oscillator interaction is
additive in projections of ${\bf r}$, the wave function is factorized, and its
nontrivial part describing scattering depends only on two variables. It
is convenient to represent these variables in the form
 $$
 x= \sqrt{ \frac{m \omega}{2}}(x_1-x_2),\ \ \ \
 y=\sqrt{ \frac{m \omega}{2}}(x_1+x_2).
 $$
The Schroedinger equation in these variables is of the form
 \begin{equation}
 \left( -\partial_{x}^{2} - \partial_{y}^{2} + x^2 +
 V(x-y)+V(x+y) -E \right)\Psi =0,
 \label{EqS}
 \end{equation}
where the energy $E$ is written in units $\omega/2$, and the potential barrier
$ V(x \pm y) = \frac{2}{\omega}V_0(( x \pm y)/\sqrt{2m\omega}  )$
is below written in the convenient form
 $
 V(X) = \frac{A}{\sqrt{2 \sigma \pi}} \exp(X^2/(2 \sigma)).
 $
Here, the amplitude $A$
is a parameter describing the energy height of the barrier, and $\sigma$
determines its spatial width. Let the process of scattering go from left to
right, and the oscillator initial state correspond to state $n$. Then
the boundary conditions are written in the form
 \begin{equation}
 \begin{array}{lr}
     \lim\limits_{y \to -\infty} \Psi    \to &
     \exp(i k_n y)\varphi_n(x) -
     \sum \limits_{j \leq N} S_{nj} \exp(-i k_j y) \varphi_j(x),
      \\
     \lim\limits_{y \to +\infty} \Psi \to &
     \sum \limits_{j \leq N} R_{nj} \exp(i k_j y) \varphi_j(x),
     \\
      \lim\limits_{x \to \pm \infty}\Psi \to &  0.
  \end{array}
 \label{BC}
 \end{equation}
The oscillator wave functions  $ \varphi_j(x)$ obey the Schroedinger equation
$$
  \left( -\partial_{x}^2  + x^2 - \varepsilon_i \right)\varphi_i =0,
$$
with energy $ \varepsilon_j = 2 j + 1 $ ($j=0,1,2,...$),
momenta $ k_j=\sqrt{E-\varepsilon_j}$,
and with $N$ being the number of the last
open channel ($ E-\varepsilon_{N+1} < 0 $).
Below, we consider an oscillator composed of bosons,
whose spectrum is conveniently numbered from 1. Thus, in what follows,
$ \varepsilon_j = 4 j -3 $ ($j=1,2,...$).

We define the probabilities of penetration $W_{ij}$ and reflection $D_{ik}$ as the
ratio of the density of a transmitted or reflected flux to that of incident
particles, i.e.
$$
W_{ij}= |R_{ij}|^2 \frac{k_j}{k_i}, \ \  D_{ij}= |S_{ij}|^2 \frac{k_j}{k_i}.
$$
It is clear that $ \sum \limits_{j \leq N} W_{ij}+D_{ij} =1.$

This problem of determination of  penetration (reflection) probabilities
requires solution of a two-dimensional differential equation. The aim of
this paper is to demonstrate the quantum transparency of a barrier.
Therefore, we take advantage of the well-known adiabatic approximation
successfully applied in various three-body problems (see, for instance,
review~\cite{AD}). To this end, we introduce basis functions  $\Phi_i$
obeying the equation
\begin{equation}
 \left( -\partial_{x}^2  + x^2 + V(x-y)+V(x+y) -\epsilon_i(y) \right)
 \Phi_i(x;y) =0,
\label{Term}
\end{equation}
and use them for the expansion  $ \Psi(x,y)= \sum \limits_i f_i(y)\Phi_i(x;y)$.
Inserting this expansion into Eq.(\ref{EqS})
and projecting onto the basis, we arrive at the system of equations
\begin{equation}
 \left( \left( -\partial_{y}^2 +\epsilon_i -E\right)\delta_{ij}
 -Q_{ij}\partial_y - \partial_y Q_{ij} + P_{ij}\right)f_j =0,
\label{Escat}
\end{equation}
where the effective interaction in channel $i$:  $E_i = \epsilon_i+ P_{ii}$
corresponds to the
diagonal part of the interaction, and functions derived in projecting
$ Q_{ij} = \langle \Phi_i, \partial_y \Phi_j  \rangle $
and $ P_{ij}=  \langle \partial_y \Phi_i, \partial_y \Phi_j  \rangle  $
correspond to the coupling of channels. Brackets mean
integration over the whole region of $x$. By definition, the functions
 $Q_{ij}$
is antisymmetric, and  $ P_{ii}$ is positive. As a rule, the coupling of
channels is small, and the processes of scattering can be described by a
limited number of equations. As in our case the spectrum of Eq.(\ref{Term}) is
discrete, a good description of the scattering processes is achieved with
the use of all the channels open in energy~ \cite{AD}. At large  $|y|$, the
effective energy $E_i \to \varepsilon_i $, and  $\Phi_i(x;y) \to \varphi_i(x) $,
which allows us to easily rewrite
the boundary conditions  (\ref{BC}) in the channel form.

Considering the boson case, we show for one channel that the effective
interaction  $E_i$ ($i=1,2$) possesses a  clear minimum, resulting in the
resonance mechanism of transparency, and that the inclusion of the second
channel does not change this picture.
\section{One-channel approximation}
\begin{figure}[ht]
\mbox{\epsfig{file=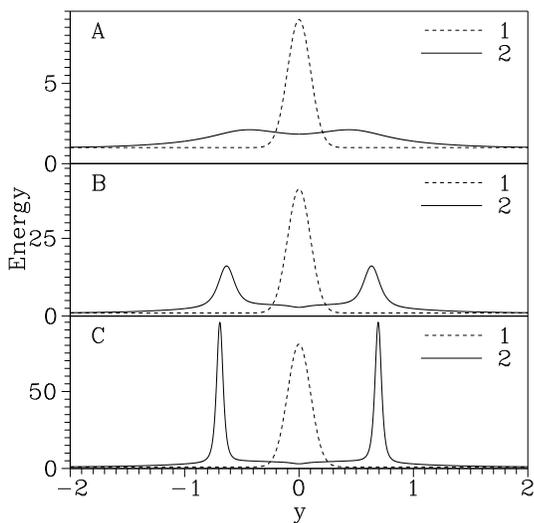,width=7cm}}
\caption {Effective energies of interaction: 1--$2V$ and 2--$E_1$. For
explanation, see the text.}
\label{Fig1}
\end{figure}

Within the chosen approach, the effect of quantum transparency is observed
even in the one-channel approximation, i.e. in the Born-Oppenheimer
approximation. In Fig.\ref{Fig1}, the dependencies $E_1(y)$ are drawn,
determined by numerical solution of Eq.(\ref{Term}) at $\sigma=0.01$
and three values of the amplitude
$A$ denoted by letters A, B, and C, respectively. These values of
parameters were taken to demonstrate the formation of a potential well
that provides the resonance peculiarities of scattering. For comparison,
shown in Fig.\ref{Fig1} are the initial potentials of barriers at
$x=0$, i.e. $ 2 V(y)$,
that describe the scattering of structureless (or extremely bound) particles.
For convenience, they are shifted by the binding energy of a pair.
%
\begin{figure}[ht]
\mbox{\epsfig{file=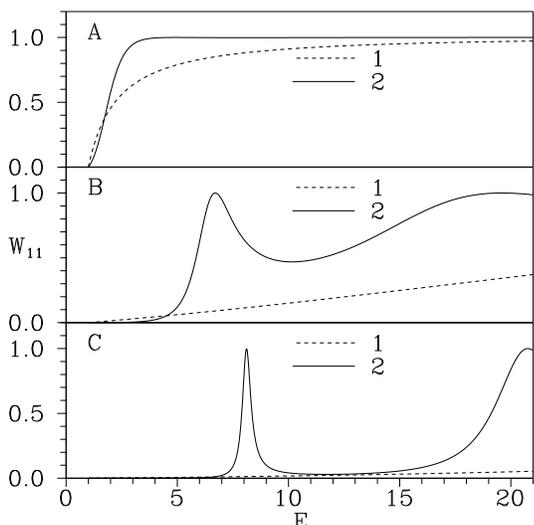,width=7cm}}
\caption {The probabilities of penetration through barriers: 1 for $2V$
 and 2 for $E_1$.  For explanation, see the text.}
\label{Fig2}
\end{figure}

In Fig.\ref{Fig2} , we present the probabilities of penetration of a pair
through barriers determined by numerical solution of Eq.(\ref{Escat})
and corresponding to the potentials drawn in Fig.\ref{Fig1}.
It is clearly seen that for $A = 1$, the
scattering of an oscillator and a structureless particle with a doubled
mass are slightly different. At $A = 5$, the resonance component of scattering
appears, and at $A = 10$, a clear resonance
with $W_{11}=1$ at the peak
is observed for energy $E_r \approx 8.12 $.
It is just this behaviour that is defined by
the term "quantum transparency of barriers". Note for comparison that
the probability of penetration through a barrier $2 V(y)$ is as few
as $\approx 0.012 $.

Complete transparency of a barrier may happen to be somewhat surprising.
Simple analogies with optical phenomena were presented in the Introduction.
Below, we write simple expressions valid in the case of rectangular
barriers and in the quasi-classical approximation which testify to the
possibility for a barrier being completely transparent. To this end, we
take the potential of form $C$ given in Fig.\ref{Fig1} with two clear peaks.
Since the one-dimensional problem of penetration through a barrier can be found
in many textbooks (see, e.g.,  \cite{Land}), here we present only the scheme of
solution of the problem of penetration through a two-peak barrier. Denoting 3
regions of classically allowed motion from left to right by numbers 1,2,3
and introducing upper indices for the amplitudes and probabilities of
penetration from the region marked by the left index to the region marked
by the right index, we easily obtain
$$
R^{(13)} = \frac{R^{(12)}R^{(23)}}{1-S^{(21)}S^{(23)}}.
$$
For simplicity, the lower index of channel 1 is omitted. Then the probability
of penetration through a two-peak channel is expressed through the
probabilities of penetration through each peak as follows:
$$
W^{(13)}=\frac{W^{(12)} W^{(23)}}{1+|S^{(21)}|^2 |S^{(23)}|^2
        - 2|S^{(21)}||S^{(23)}| \cos(\theta) },
$$
where $\theta $  is a doubled difference of phases (or action in quasi-classics)
of motion between the left and right peaks. Time-reversal invariance leads
to the principle of detailed balance (see, e.g. \cite{Land})
that in our case results in the equality $|S^{(21)}| =  |S^{(12)}|$.

For a symmetric potential ($ W^{(12)} = W^{(23)}$), the penetration probability
 $ W^{(13)}$ reaches
a maximum at $\theta = 2 \pi$n (n=1,2,...).
Note that it is just the condition that in a quasi-classical approach
determines the spectrum of bound states for infinitely
broad peaks. Provided that  $ |S^{(ij)}|^2=1-W^{(ij)} $,
it is not difficult to verifies that at
these energies,  $ W^{(13)}=1$. i.e. a complete transparency occurs.

Parameters of the barrier potential $V$ were chosen so that the resonance
energy  $E_r$ would be higher than the energy of the second channel $\varepsilon_2 = 5 $.
This is necessary for proving that the inclusion of inelastic processes
does not change the resonance picture of transparency.

It is necessary to note, that the parameters of a potential V differ
from parameters of a potential of Ref.\cite{Rub1},
where power height of a barrier is comparable to energy of elementary
excitation ($ V (0) = 1, \ \ \omega = 1/2 $).
The parameters of potentials will become close if magnitude $ \omega $
from Ref.\cite{Rub1} will be in 50 times less.
\section{Two-channel approximation}
\begin{figure}[ht]
\mbox{\epsfig{file=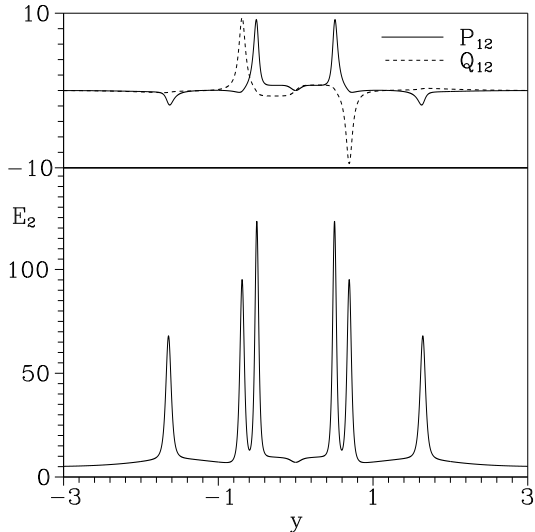,width=7cm}}
\caption
{
Components of the second channel. For explanation, see the text.
}
\label{Fig3}
\end{figure}

In Fig.\ref{Fig3}, we show the results of numerical solution of Eq.(\ref{Term})
for the second channel.
It is seen that the coupling functions of channels $Q_{12}$ and $P_{12}$
are about 2 ordered as small as diagonal values  $E_2$. The
effective energy  $E_2$ is more complicated in form than  $E_1$ and can also
generate extra resonances, whose correct consideration requires inclusion
of the third channel (energies above 9). This goes beyond the scope of
our problem of demonstrating the transparency of a barrier, and here we
only mention the presence of peaks of  $E_1$ in  $E_2$.
%
\begin{figure}[ht]
\mbox{\epsfig{file=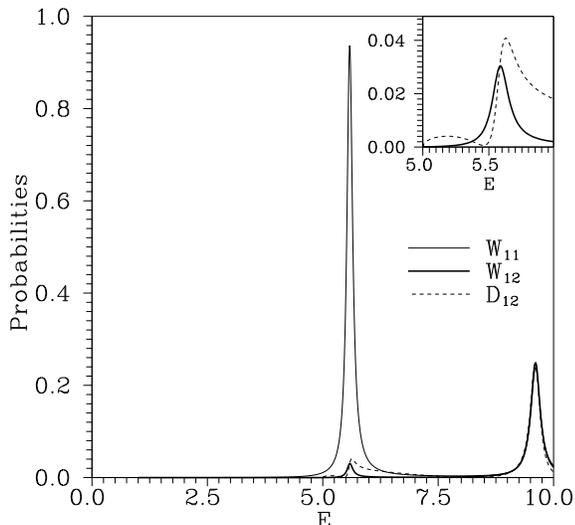,width=7.5cm}}
\caption{
Probabilities of penetration through barriers. For explanation, see the text.
}
\label{Fig4}
\end{figure}

In Fig. \ref{Fig4}, we plot the probabilities of penetration through
a barrier of an oscillator in the ground state.
The elastic peak  $ W_{11}$ is conserved though
has a shift ($E_r \approx 5.58$) and a considerable smaller width, about 3 times.
Its maximal value $ \approx 0.94 $ does not reach 1 owing to the
opened second channel.
The quantities $W_{12}$ and $ D_{12}$ shown in the region of resonance energies
at the top of the Figure (insertion) amount to  $ \approx 0.03$ and the total
probability of penetration through a barrier reaches  $\approx 0.97$,
which allows us to speak about a considerable, though not 100\%, transparency.
Note that the probability of penetration of the barrier  $ 2 V(y)$ in this
region is only  $ \approx 0.0075$. The quantity $D_{11}$ is very closed to zero
($\approx 0.0007$), thus demonstrating the above optical effect
of suppression of a reflected wave even in the two channel case.

The second peak at energy  $E_r \approx 9.6$ shown in Fig.\ref{Fig4}
cannot be considered reliable since at these energies, account is to
be made of the third channel.
This peak demonstrates as interesting peculiarity -- all the
probabilities of both channel 1 and channel 2 amount to 1/4. Moreover,
the behaviour of probabilities of transition from state 2
( $W_{22}, W_{21}, D_{21}$) in the energy
region of the second peak (not shown in Fig.\ref{Fig4}) is visually
nondistinguishable from the behaviour of inelastic components from state 1.
Owing to the principle of detailed balance  \cite{Land}, the equalities
$  W_{21} = W_{12} $ and $ D_{21} = D_{12}$
demonstrate only the accuracy of calculation.  Of surprise is the
behaviour of the probability  $ W_{22}$ whose energy dependence with 5\% accuracy
reproduces the behaviour of inelastic components around this peak.
\section{Conclusion}
The considered mechanism of transparency of barriers for a coupled pair of
particles is clearly observed for narrow and high barriers, as compared to
characteristic sizes and energy of an oscillator. These conditions do not
eliminate spatially asymmetric barrier potentials since the symmetry of
effective interaction  $E_i(y)$ is determined only by particles being identical.
Therefore, the effects of quantum transparency can occur in various fields
of physics. Here note is to be made that when it is allowable
to describe the processes of the false vacuum  disintegration
in high energy collisions by means
of quantum tunnelling of a pair of particles coupled by an oscillator
interaction, there arise mechanisms of the resonance transparency of a
barrier, much increasing the decay probabilities of the false vacuum.

\bigskip
{\bf Acknowledgements.}
The author is indebted to V.A.Rubakov,
which has marked not full adequacy of the considered quantum mechanical system
to a problem of induced disintegration of false vacuum
for a very useful discussions.

\end{multicols}

\begin{thebibliography}{99}
\bibitem{Zakh}   B.N. Zakhariev and A.A.Suzko {\it Direct and Inverse Problems,
Potentials in quantum Scattering} (Springer-Verlag, Berlin Heidelberg, 1990).
\bibitem{Rub1} G.F. Bonini, A.G. Cohen, C. Rebbi, V.A. Rubakov,
Phys.Rev. {\bf D60},  076004 (1999), (e-Print Archive: hep-ph/9901226).
\bibitem{Rub2} V.A. Rubakov, M.E. Shaposhnikov,
Usp.Fiz.Nauk { \bf 166}, 493 (1996) (in Russian),
(e-Print Archive: hep-ph/9603208)
\bibitem{Tyn1} A.N. Kuznetsov, P.G. Tinyakov,
Phys.Rev. { \bf D56}, 1156 (1997),
(e-Print Archive: hep-ph/9703256).
\bibitem{AD}     S.I.Vinitsky and L.I.Ponomarev, Sov.J.Part.Nucl. {\bf 13}, 557 (1982).
\bibitem{Land}   L.D.Landau and E.M.Lifshitz, {\it Quantum Mechanics: Non-relativistic
Theory} ( 3rd ed., Pergamon Press, Oxford, 1977).
\end{thebibliography}
\end{document}